\documentstyle [12pt] {article}

\catcode`@=12
\begin{document}
\vspace{0.5in}
\oddsidemargin -.375in  
\newcount\sectionnumber 
\sectionnumber=0 
\def\be{\begin{equation}} 
\def\ee{\end{equation}}
\thispagestyle{empty}
\begin{flushright} AMES-HET-96-02\\April 1996\\
\end{flushright}
\vspace {.5in}
\begin{center}
{\Large \bf Implications Of A Non-Standard Light Higgs Boson \\}
\vspace{.5in}
{\rm A. Datta, B.L. Young and X. Zhang \\}
{\it Department of Physics and Astronomy \\}
{\it  Iowa State University, \\}
{\it Ames, Iowa
 50011 \\}
\vskip .5in
\end{center}  
\begin{abstract}
 Analyses of the vacuum stability of the electroweak
theory 
indicate that
 new physics occur
at a scale of order of 1 TeV if a light Higgs is discovered at LEP II. 
In this paper, we parameterize the
effects of new physics in the effective Lagrangian approach and examine
its implication on the Higgs boson production at LEP II. We
consider the effect of higher dimension  
 operator on the Higgs potential and
 calculate the lower
bound on the Higgs boson mass from the requirement of vacuum stability.
We show that if a Higgs boson is seen at LEP II then under favourable
conditions the
deviation of the production cross section from the standard model value
could be significant and therefore the presence of the new physics is
detectable at LEP II.
\end {abstract}
\newpage
\baselineskip 24pt

\section{\bf Introduction}

  One of the important issues in particle physics is to
understand the origin of the electroweak scale. In the standard model, the
electroweak symmetry breaking arises from a complex
fundamental Higgs scalar. However, the theoretical arguments
of "triviality" \cite{1} and "naturalness" \cite {2}, 
suggest that such a simple spontaneous
symmetry breaking mechanism may 
not be the whole story. This leads to the belief 
 that the Higgs sector of the
standard model is an effective theory. The advent of new physics
which reveals a more fundamental structure underlying the symmetry breaking
 mechanism can be
defined by an energy scale $\Lambda$ which also serves as a cutoff
of the effective theory. 

As a  requirement of the Higgs sector, the effective potential
should have 
 a global minimum at the electroweak scale ($ v=246 GeV$). This is the
condition of vacuum stability\cite{Sher}. 
 For the existence of a light Higgs 
boson, within the energy
range of LEP II, the standard model 
vacuum will become unstable at the order 
of 1 TeV, because of the large destabilizing effect of the top quark
contribution to the effective potential.
This can be taken as an indication of the presence
of new physics around this scale. With such a low cutoff 
one expects effects of new physics to show up relatively soon, even in 
 experiments at LEP II.

There are various proposals of new physics beyond the standard model,
such as SUSY, Left-Right models, multi-Higgs models,
composite Higgs models, Top quark condensation models, etc.
Recently Hung and Sher \cite{HS}  has studied
 the question of vacuum stability 
in a specific model with a singlet scalar added
 to the standard model. In this paper we
consider a model independent approach to new physics, {\it i.e.}, the 
effective Lagrangian approach, and analyze the implication of
 vacuum stability
 on the Higgs boson mass and its production. Our analysis 
show that, if the Higgs
boson is discovered at LEP II then the scale of new physics should be around 
O(1 TeV), and 
the new physics effects on 
the Higgs production can be sizable. For
  instance, for a Higgs of mass of
75 GeV, the correction to the cross section for
$e^{+} e^{-} \rightarrow Z H $ due to new physics can be
  around $8 -11 \%$, which is detectable at LEP II.

This paper is organized as follows: in Section 2, we analyze the Higgs
boson
 mass
bound in the effective theory. In Section 3, we discuss
 the possible effects of new
physics on Higgs boson production at LEP II and in Section 4, we
  briefly summarize our results.

\section{\bf Effective theory and Higgs mass bound }

 In the  effective Lagrangian approach to new physics, the leading 
terms are given by the
standard model. The corrections which come from a certain underlying
theory
beyond the standard model are described by higher dimension
operators,
\begin{eqnarray}
{\cal L}^{new}  =  \sum_{i} \frac{c_i}{\Lambda^{d_i-4}} {\cal O}^i , \ 
\end{eqnarray}
 where $d_i$ are the dimensions of ${\cal O}^i$, which are
 integers greater than 4. The operators ${\cal O}^i$ are
$SU(3)_c \times SU(2)_L \times U(1)_Y$ invariant and
 contain only the standard model fields. The dimensionless
 parameters $c_i$,
determining the strength of the contribution of operators ${\cal O}^i$,
can be calculated by matching the effective theory with the underlying
theory. In general, if the new physics is due to
 a strongly interacting system, for instance in
a composite Higgs model \cite{dh} or with a low scale top condensate models
\cite{tc}, $c_i$ are expected to be of O(1). 
       For weakly coupled new
physics the parameters $c_i$ may be an order of magnitude smaller.

 Analyses of higher dimension
operators
have been performed by
many authors \cite{bu}. In
this paper we consider only CP conserving operators which 
can be constructed out of
the Higgs fields $\Phi$, covariant derivatives of the Higgs field, $D_\mu
\Phi$,
and the field strength tensors $W_{\mu\nu}$ and
$B_{\mu\nu}$ of the $SU(2)$ and the $U(1)$ gauge fields. There are
 8
dimension-six operators denoted by ${\cal O}_{\Phi, 1}, 
{\cal O}_{\Phi, 2},
{\cal O}_{BW}, {\cal O}_{W}, 
{\cal O}_{B}, {\cal O}_{WW}$, ${\cal O}_{BB}$
 and ${\cal O}_{\Phi, 3}$. They modify the standard model Lagrangian to
the $1/\Lambda^2$ order and
the effective
Lagrangian for new physics, ${\cal L}^{new}$, up to dimension 6, is 
 given by
\begin{eqnarray}
{\cal L}^{new} & = & \frac{1}{\Lambda^2}
 c_{\Phi, 3} {( \Phi^+ \Phi - {\frac{v^2}{2}} )}^3 \nonumber\\
                 & + & \frac{1}{\Lambda^2}\left[
 c_{\Phi, 1} 
{( D_\mu \Phi )}^+ \Phi \Phi^+ ( D^\mu \Phi )
                   + \frac{1}{2}c_{\Phi, 2} \partial_\mu( \Phi^+ \Phi )
                       \partial^\mu ( \Phi^+ \Phi )\right]  \nonumber\\
  & + & \frac{1}{\Lambda^2}\left[ c_{BW} 
\Phi^+ \hat{B_{\mu\nu}}\hat{W^{\mu\nu}} \Phi
                   + c_W {( D_\mu \Phi )}^+ \hat{W^{\mu\nu}} ( D_\nu
\Phi )\right] \nonumber\\
   & + & \frac{1}{\Lambda^2}\left[ c_B 
{( D_\mu \Phi )}^+ \hat{B^{\mu\nu}}(D_\nu  \Phi )
+ c_{WW} \Phi^+ \hat{W_{\mu\nu}}\hat{W^{\mu\nu}} \Phi\right]\nonumber\\
               & +  & \frac{1}{\Lambda^2} c_{BB}
\Phi^+ \hat{B_{\mu\nu}}\hat{B^{\mu\nu}} \Phi. \
\end{eqnarray}
 Only ${\cal O}_{\Phi, 3}$
  contributes to the effective Higgs potential. All the other operators
except $c_{\Phi,3}$ will
 contribute to the Higgs boson production and
in Section 3 we will study the effect of these operators in Higgs boson
 production.
 
In the presence
of the higher dimensional operator ${\cal O}_{\Phi, 3}$ the
 tree level Higgs potential can
now be written as
\begin{eqnarray}
V_{tree} & = & -\frac{m^{2}}{2}\phi^{2} +\frac{1}{4}\lambda \phi^{4} +
\frac{1}{8}\frac{c_{\phi,3}}{\Lambda^{2}}
{(\phi^2 -v^2)}^3, \
\end{eqnarray}
which is corrected by the one-loop term, $V_{1loop}$,
\begin{eqnarray}
V_{1loop}(\mu) & = & \sum_{i} \frac{n_i}{64 \pi^{2}}M_i^4(\phi)
\left[\log{\frac{M_i^2(\phi)}{\mu^2}}
-C_i\right], \
\end{eqnarray}
where
\begin{eqnarray}
M_i^2(\phi) & = & k_i\phi^2 -k_i' \nonumber.\\
\end{eqnarray}
The summation  goes over the gauge bosons, the
fermions and the scalars of the standard model.
 The values of the
constants $n_i$, $k_i$, $k_i'$ 
and $C_i$ can be found in Refs\cite{Sher,Casas}.
The one-loop effective potential, including the higher 
dimensional operators, is
$$ V = V_{tree} + V_{1loop}. $$


In Fig.(1) we plot the effective potential for $\Lambda=4$ TeV and
for three typical values of $c_{\phi,3}$. We see that the effect of a positive
$c_{\phi,3}$ is to delay the onset of vacuum instability compared to
the standard model while the effect of a negative
$c_{\phi,3}$ is to accelerate the onset of vacuum instability.

To obtain a lower bound on the Higgs boson mass,
 in the absence of higher dimensional operators one
can take the location of vacuum instability to be as large as $\Lambda$.
However, in our approach, for the low energy theory
to make sense, we should require $\phi < \Lambda$. We
 take the scale of vacuum instability, $\Lambda'$, to be $0.5\Lambda$, so
 the corrections from operators of dimension greater than six to our
result is suppressed by a factor of  
$\frac{{\Lambda'}^2}{{\Lambda}^2} =0.25$. In
 Fig. (2) we plot the lower bound of the Higgs
mass versus $\Lambda$.


 Since we are dealing with values of the field
$\phi$ larger than $v$, we need to consider a renormalization group 
improved potential
for our analysis \cite{Sher,Casas,L,LS,S,Al}.
 Working
with the one-loop effective potential, we
  consider two-loop
running for $\lambda$, the top Yukawa coupling ($g_Y$), gauge couplings
and the Higgs mass. This procedure resums all next-to-leading 
logarithm contributions \cite {Kas}.
 The various $\beta$ functions to two-loop order
 can be found in Ref
\cite{ME}. 
After obtaining the running Higgs boson mass, the physical pole mass
\footnote{
The effect of the operator $c_{\Phi,2}$ in Eq. (2) causes an finite
renormalization of the Higgs field H\cite{bu}.
$ H \rightarrow Z_H^{\frac{1}{2}} H, $ where
$ Z_H^{-1} = 1+c_{\Phi,2}\frac{v^2}{\Lambda^2}$. This gives rise to
a correction to 
the
Higgs boson mass by $\sim 2 \% $ for
$c_{\Phi, 2} \sim O(1)$. }           
 can be calculated.
The relevant
equation relating the running mass to the pole mass
 can be found in Ref \cite{Casas}.
The boundary conditions for the gauge couplings and the top quark
Yukawa couplings are known at the electroweak scale in terms of the
measured values, taking into account
the connection between the running top mass
 and the pole top mass measured at
$175$ GeV. The vacuum stability requirement
provides the boundary condition for $\lambda$ at the scale
 $\Lambda'$, which is given by
\begin{eqnarray}
\lambda_{eff}(\Lambda') &\approx  &-\sum_{i} \frac{n_i}{16
\pi^{2}}{k_i}^2(\log{k_i}-C_i) -\frac{1}{2}\frac{\Lambda'^2}{\Lambda^2}
c_{\Phi,3} ,
\end{eqnarray}
where
\begin{eqnarray}
\lambda_{eff}(\Lambda') & = & \lambda(\Lambda')
 -\frac{3}{2}c_{\Phi,3}\frac{v^2}{\Lambda^2} . \
\end{eqnarray}
In Fig. (3) we show the renormalization 
group improved Higgs mass bound versus $\Lambda$. We
 see that for a light Higgs mass within the discovery range of LEPII, the
new physics scale is a few TeV. For $c_{\Phi,3}=1.0$ the
higher dimensional operator helps to significantly stabilize the vacuum
to such an extent that the scale of new physics will be too large to show
any significant effect at LEP II.

\section{\bf Higgs Boson Production}

 In Section 2, we have considered the Higgs 
mass bound that stabilize the
electroweak vacuum in the effective theory.
 Turning the argument around we can see from Fig.
(3) that if the Higgs boson is found at LEP II then we
can read off the upper bound for the scale of new physics. As an example, with
 $c_{\Phi,3}=-1.0$ and $m_H=75$ GeV the new physics
scale is $\stackrel{<}{\sim} 1.08$ TeV.
This is perhaps as much as can be said about the effect of the operator
$O_{\Phi,3}$. However the presence of the other operators in Eq .(2) will
modify the couplings of the Higgs boson interactions. The determination
of a low scale for new physics will allow the effects of the other
operators to manifest themselves more readily. In this section 
we will show that
under favourable conditions new physics effects can be visible at LEP II
and the mechanism of new physics, strong versus weakly interacting may
also be discernible. 
We 
expect this new physics effect to manifest in the
 Higgs production, $e^+e^-\rightarrow Z H$
{\footnote{ This process has
recently received attention as a probe for new physics at LEP II
\cite{WU}.}}
at LEP II.
The
Lagrangian given in Eq .(2) gives rise to
anomalous Higgs couplings, which can affect Higgs boson production.
Following
Ref\cite{bu} we write down the relevant vertices generated 
from the Lagrangian in
Eq. (2) in terms of the physical Higgs field, $H$.
\begin{eqnarray}
{\cal L}_{eff} &= & \frac{g M_W}{\Lambda^2}\left[
T_1 H Z_{\mu}Z^{\mu} +T_2 Z_{\mu \nu}Z^{\mu}(\partial^{\nu}H)
+T_3 H Z_{\mu\nu}Z^{\mu\nu}
+T_4 A_{\mu \nu}Z^{\mu}(\partial^{\nu}H)
+T_5 H A_{\mu\nu}Z^{\mu\nu}\right],\
\end{eqnarray}
with
\begin{eqnarray*}
T_1 &= &\frac{2m_W^2}{g^2}\frac{c_{\Phi,1}}{c^2},\nonumber\\
T_2 &= &\frac{c^2 c_{W} +s^2c_B}{2c^2},\nonumber\\
T_3 &=&-\frac{c^4 c_{WW} +s^4c_{BB} +s^2c^2c_{BW}}{2c^2},\nonumber\\
T_4 &=&\frac{s(c_W-c_B)}{2c},\nonumber\\
T_5 &=&\frac{s(-2c^2 c_{WW} +2s^2c_{BB} +(c^2-s^2)c_{BW})}{2c},\nonumber\\
\end{eqnarray*}
where $g^2=e^2/s^2=8m_W^2G_F/\sqrt{2}$ and s and c are the sine and
 cosine of the Weinberg angle.. We also include the
 effects of the Higgs boson wavefunction renormalization due to
operators
$O_{\Phi,1}$ and $O_{\Phi,2}$,
\begin{eqnarray}
{\cal L}_{ren} & = & \frac{1}{2}g_{Z}M_{Z}
\left[1-\frac{c_{\Phi,1}+c_{\Phi,2}}{2}
\frac{v^2}{\Lambda^2}\right]HZ_{\mu}Z^{\mu}.\
\end{eqnarray}
The strength of the anomalous Higgs couplings depends on the values
of various coefficients $c_i$. In a strongly interacting theory
for the Higgs sector, such as composite Higgs boson
models\cite{dh} and low scale top condensation models\cite{tc},
it is  difficult to calculate the absolute values of $c_i$. However, one
expects in general that $c_i \sim O(1)$
 {\footnote{ Operator ${\cal O}_{\Phi, 1}$ 
contributes to the $\rho$ parameter and is therefore tightly
constrained. We set $c_{\Phi,1}=0$ in our calculation by assuming
the existence of a custodial 
$SU(2)$ in these models. The coefficients of the other operators are
taken to be $O(1)$. 
 The operator ${\cal O}_{BW}$ contributes to the S
parameter of Peskin and Takeuchi. For our choice of the new physics scale given
above the correction of the operator ${\cal O}_{BW}$ to S is $\sim -0.6$
which is within the experimental limit on S \cite {Hewett}.}}. In 
Fig. (4) we show the cross sections for the process
$e^+e^-\rightarrow Z H$ for $c_i\sim 0(1)$. The formula 
for the cross sections can be found
in Ref\cite{WU}.

 We see that new physics effects are at a level
of 
$$R=\frac{ \sigma_{NSM}-\sigma_{SM}}{\sigma_{SM}}=8-11 \% ,$$
where $\sigma_{SM}$ is the standard model cross section and
$\sigma_{NSM}$ is the cross section with the inclusion of anomalous couplings 
\footnote{ We have not included radiative corrections to the cross
section
because they are known to be small at LEP energies \cite {HA} and the
percentage change to cross
section with or without anomalous couplings 
due to radiative corrections will approximately be equal. The ratio
$R$, therefore, will remain almost the same with or without
 radiative corrections.}.
 In LEP II with a center of mass energy 175-205 GeV and 
  an integrated luminosity of $300-500 pb^{-1}$ new physics effects 
on the Higgs boson production with
the magnitude mentioned above
 will be detectable\cite{willen}. However if the new physics is weakly
interacting,
$c_i$ may be of the order of $0.1$ or smaller, then the correction 
to the Higgs production cross section will be
too small to be visible at LEP II.
            
\section{Summary}
 In summary, we have re-examined the 
Higgs mass bound from the requirement of vacuum
stability in the effective theory by
taking into account the contribution of higher dimension
operators to the effective potential\footnote{
High dimension operator correction to the effective 
potential for electroweak baryongenesis was considered
in \cite{zhang}.}. We
show that if the Higgs boson is discovered at LEP II,
new physics could be around TeV and
its effect on Higgs boson production 
will be observable at LEP II in certain models with strong interaction as
the underlying dynamics of the Higgs sector.
             
             
{\bf Acknowledgment:} We would like to thank
P.Q. Hung, 
G. Valencia and
D. Zeppenfeld for discussions. This work was supported in part
by DOE contract number DE-FG02-92ER40730(A. Datta) and DE-FG02-94ER4-817 
(B.-L. Young and X. Zhang).

\newpage

\section{\bf Figure Captions}
\begin{itemize}
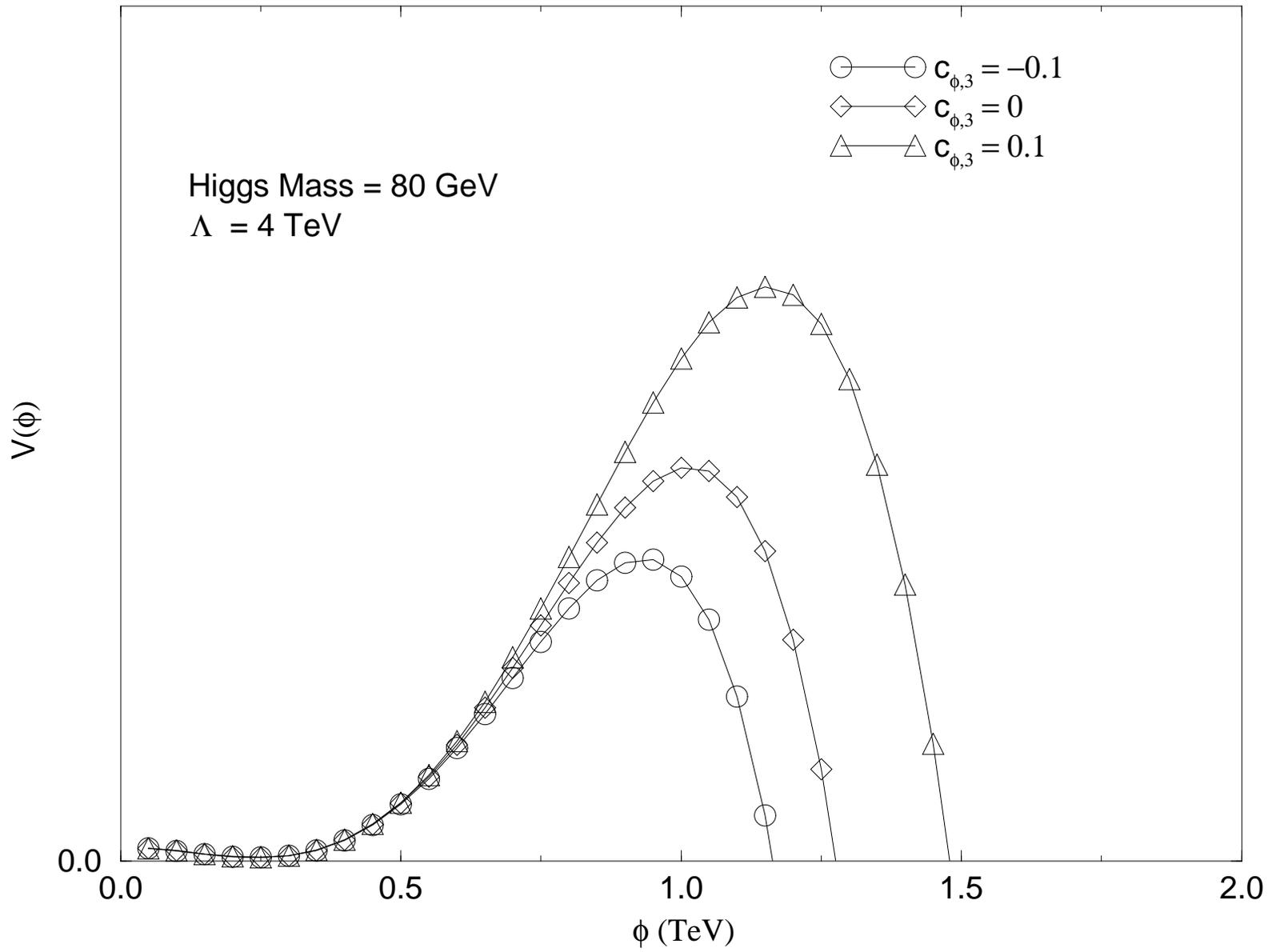
\item[\bf{Fig. 1}] The effective potential for various values of
$c_{\phi, 3} $. The Higgs mass is taken as $80 GeV$ and the scale of 
new physics $\Lambda =4 TeV$. The curve with $c_{\phi,3} =0$ corresponds
to
the standard model.

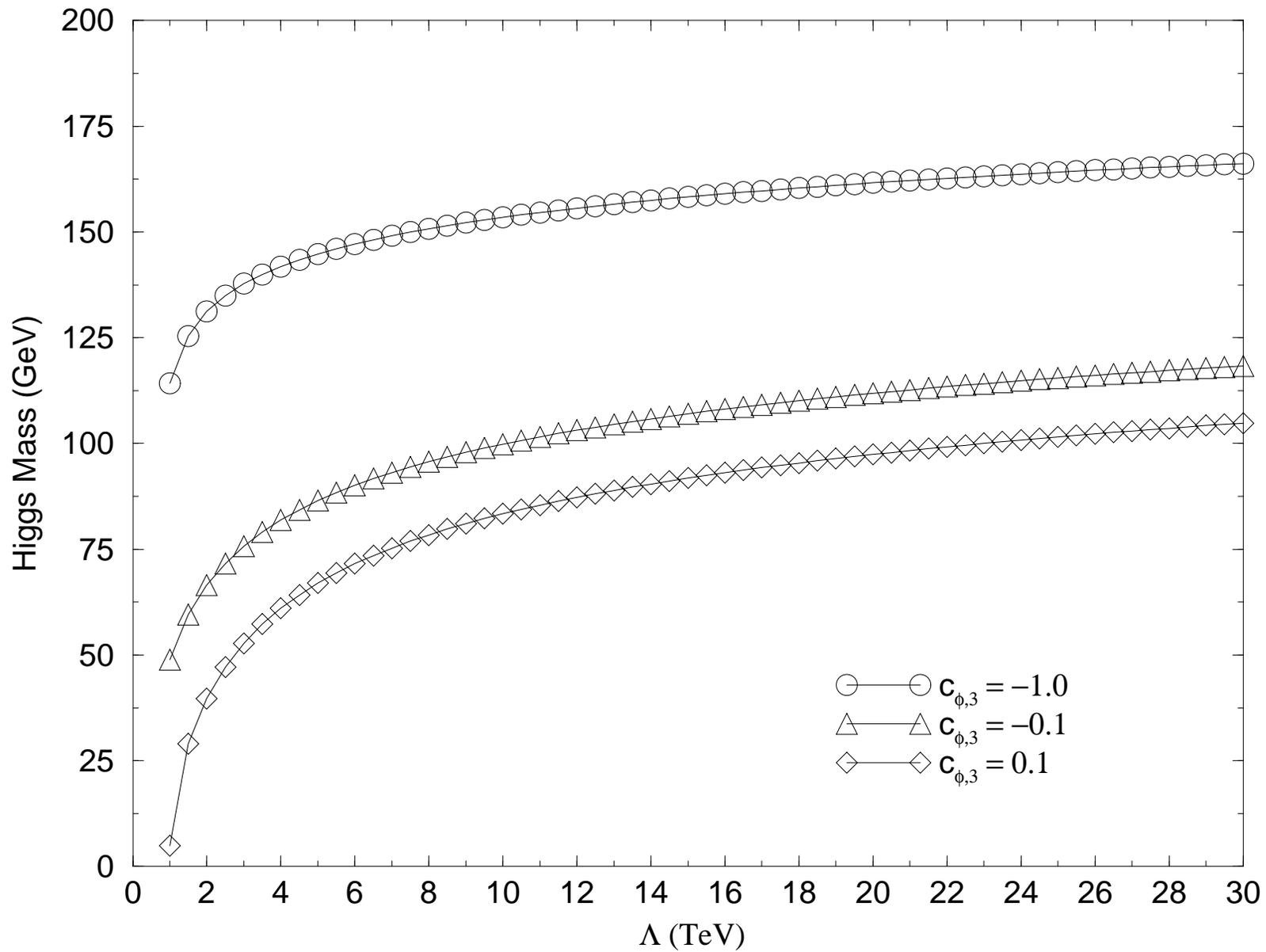
\item[\bf{Fig. 2}] The lower bound on the Higgs mass as a function of
the new physics scale $\Lambda$. The scale of vacuum stability has been
chosen to be $\Lambda'= \frac{\Lambda}{2}$. The scale in the effective
potential is set at $\mu=v=246 GeV$.

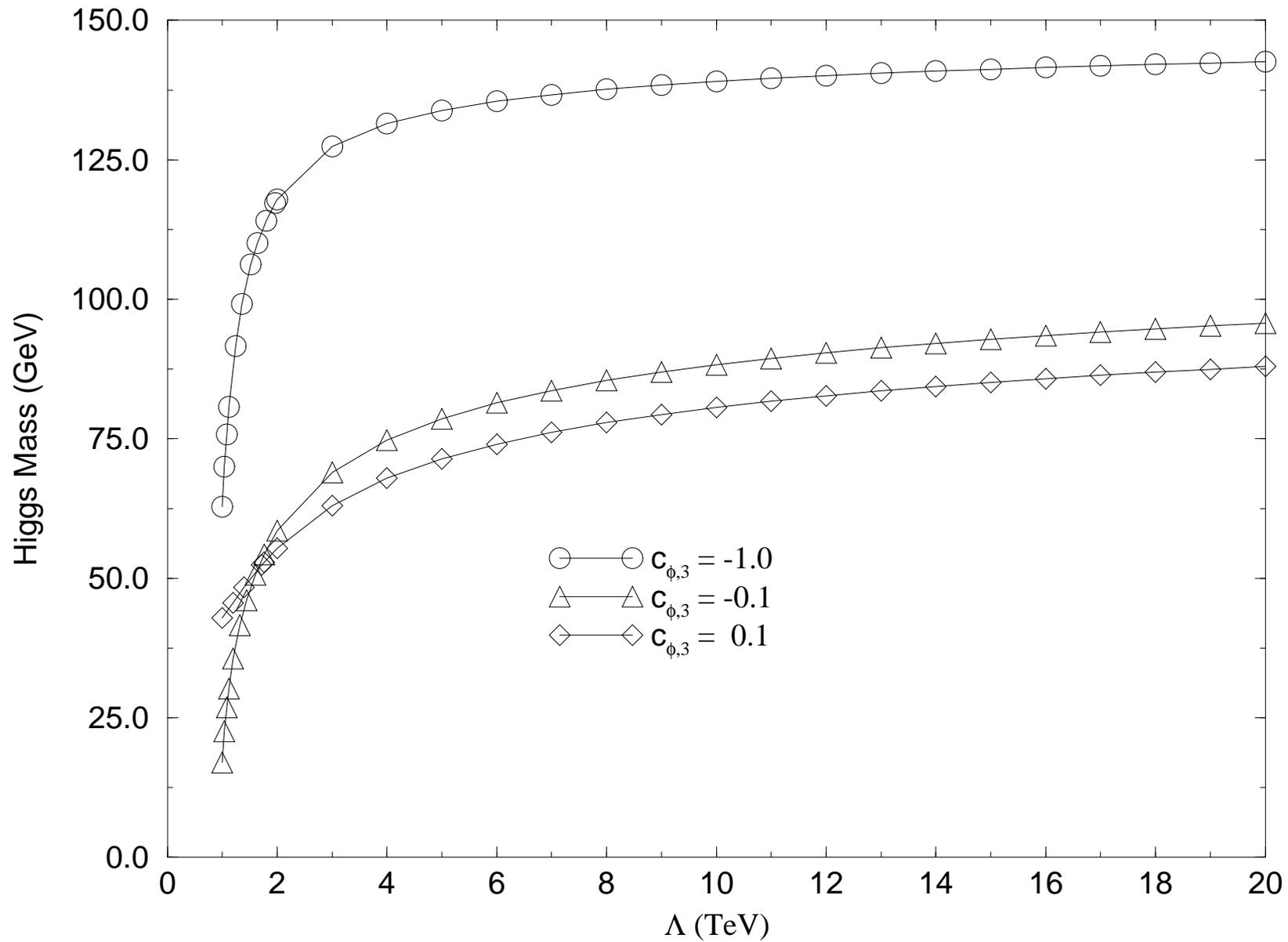
\item[\bf{Fig. 3}] The lower bound on the Higgs mass as a function of
the new physics scale $\Lambda$ for various $c_{\phi, 3}$. The scale 
of vacuum stability has been
chosen to be $\Lambda'= \frac{\Lambda}{2}$. A renormalization group
improved
effective potential has been used and the running of the couplings in
the potential have been considered up to two loop order.

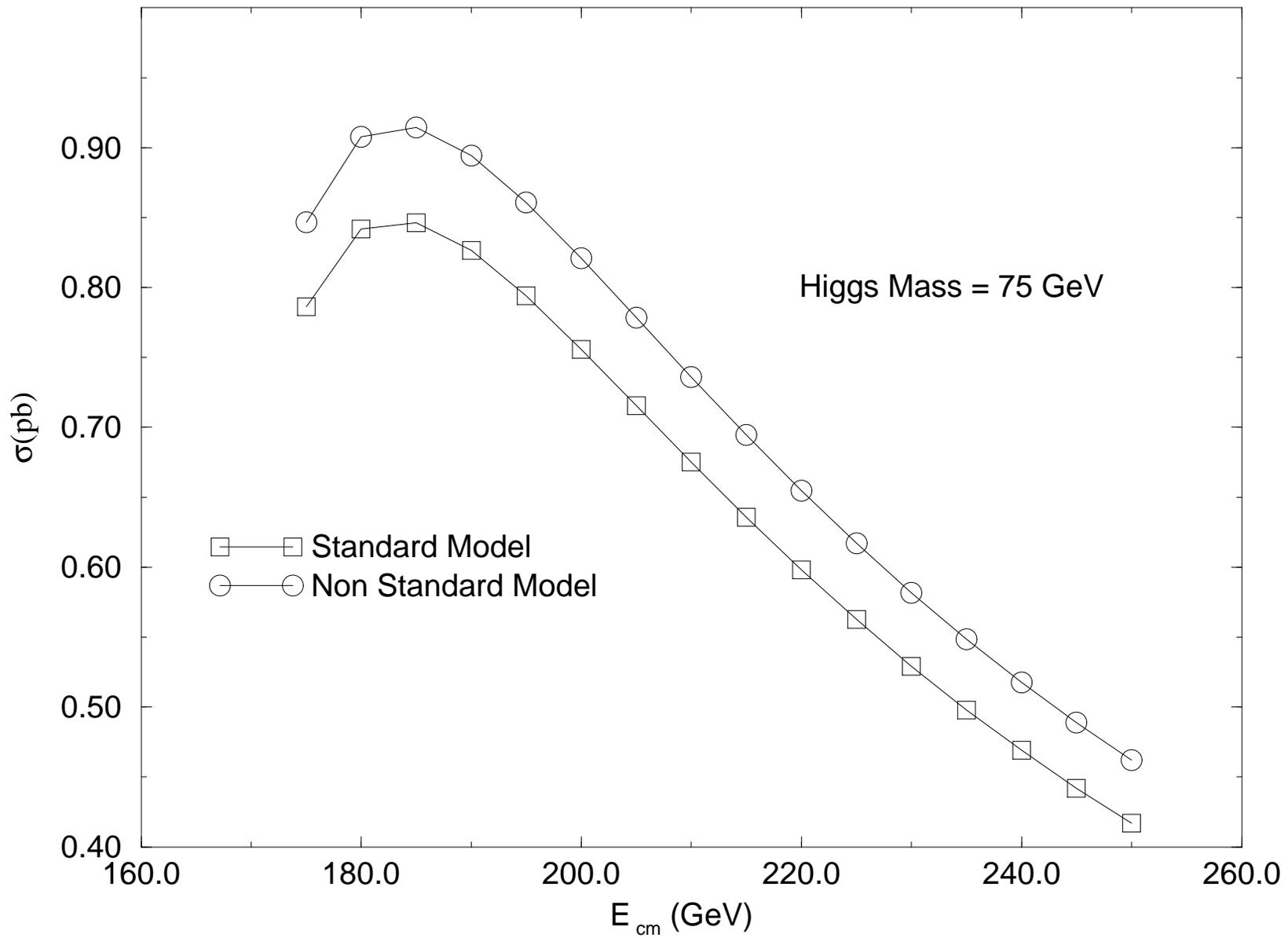
\item[\bf{Fig. 4}] The cross section for $e^+ e^{-} \rightarrow Z H$ for
a Higgs mass of $75 GeV$ in
the standard model and including anomalous couplings.
 
\end{itemize}

\end{document}